
\magnification=\magstep1
\centerline{STAR-FORMATION KNOTS IN IRAS GALAXIES}
\vskip 10pt
\centerline{J. B. Hutchings\footnote{$^1$}{Guest Observer, Canada France Hawaii
Telescope, which is operated by NRC of Canada, CNRS of France, and the
University
of Hawaii}}
\centerline{Dominion Astrophysical Observatory, National
Research Council of Canada}
\centerline{5071 W. Saanich Rd, Victoria, B.C. V8X 4M6, Canada}
\vskip 20pt
\centerline{To appear in the Astronomical Journal}
\vskip 20pt
\centerline{ABSTRACT}
\vskip 10pt
Images of IRAS galaxies with a range of IR properties are examined for
bright knots, both within and outside the galaxy. These are found almost
exclusively in galaxies with steep IR spectra, but over a wide range of
IR luminosity, and usually without strong nuclear activity. In most cases,
the knots are likely to
be star-formation induced by tidal interactions, and are seen in the
early stages of such interactions. Detailed photometry is presented of knots in
six
representative galaxies. The knots appear to have a wide range of colour
and luminosity, but it is argued that many are heavily reddened. Knots formed
outside the parent galaxy may be a new generation of what later
become globular clusters, but they appear to have a wide range of luminosities.
\vfill\eject
\centerline{1. INTRODUCTION}
\vskip 10pt
  Hutchings and Neff (1991:HN) published a discussion of the morphology
and colours of a sample of 64 IRAS point sources. Their sample was chosen to
cover the widest possible range of 60$\mu$ luminosities and 25/60$\mu$
ratio (the IR `H-R diagram'), and includes objects ranging from local
irregulars to QSOs, Seyfert galaxies, and superluminous interacting galaxies.
The data were images in B and R taken with the Canada France Hawaii
telescope, with image FWHM in the range 0.7 to 0.9 arcsec. Most of the
galaxies are clearly in a state of tidal interaction or merging, and HN made
a numerical assessment of the strength and `age' of the tidal events for all
the objects. HN also note that the IR-diagram segregates many of the galaxies
by spectroscopic types, and discuss how they may represent time sequences
initiated by interactions between galaxies. In their Table 2, they give brief
notes on the galaxy images, and in particular mention that 17 of the 64 have
bright
knots in them.

   It has since become clear that interactions appear to give rise to
widespread star-formation in many galaxies, not necessarily selected only by
high IR flux. It has also been noted that the brighter knots in star-forming
galaxies have similar properties (Holtzman et al 1992, Richer et al 1993,
Hunter, O'Connell and Gallagher 1994, O'Connell, Gallagher and Hunter 1994,
Crabtree and Smecker-Hane 1994).Thus they may represent an upper luminosity or
size limit for young star-formation in a cluster.
If so, these objects may be used either as standard candles, or as
measures of abundance or IMF in distant environments.

A lower luminosity
example of such an object in the local environment is 30 Dor in LMC.
The FUV imaging of nearby galaxies by the UIT telescope
on the ASTRO missions, has shown that there are many knots like these
that have high UV flux, and hence presumably contain very young massive
stars. UV spectroscopy of such knots in the circumnuclear region of
the Seyfert galaxy NGC 1068 shows that the UV spectrum can be modelled
in this way, and conclusions drawn on their age and IMF (e.g. Hutchings et al
1991)

    It has further been speculated in the above references
that these blue clusters may be
the early forms of globular clusters. Spectroscopic studies are under way
to observe and model the spectra and deduce the ages of such clusters. If
globular clusters can form during galaxy encounters, then the globular
cluster population of a galaxy may contain information on its merging
and star-forming history as well as its initial formation.

   It thus seems clear that the sample of IRAS galaxies of HN form a useful
database for studying some of these questions, being large, of uniform
quality, and not selected by the presence of bright knots or interaction.
There is associated IR information and existing study of interaction status,
which appears to be a relevant parameter in the formation of the knots.
In addition, many of the objects were also studied at radio wavelengths by
Neff and Hutchings (1992: NH).

    In this paper, the CCD images in the HN sample are examined for bright
knots, and the
galaxies are classified by the appearance of the knots. Connections with
interaction status, other morphology, and spectroscopic type are examined.
Further, detailed 2-colour photometry is presented and discussed
on a representative sample of the galaxies.
\vfill\eject
\centerline{2. BRIGHT KNOTS IN IRAS GALAXIES}
\vskip 10pt
   In Table 2 of HN, 17 of the sample are noted as having bright knots.
There is a wide range of redshift in the sample, and clearly the visibility
of small knots is a function of redshift. Indeed, knots are noted only
in objects of redshift $<$0.040. If the parent sample is restricted to this
redshift range, there are 17 with noted knots and 23 without. Table 1
compares the properties of these two groups, from the values in HN Table 2.
Figure 1 shows where different galaxy subgroups lie in the IR-diagram.

    There are several things to note from Table 1 and Figure 1. The
elimination of objects with redshift higher than 0.040 cuts out principally
the high luminosity, flat IR-spectrum sources (many are QSOs or Seyfert 1s).
The knotty galaxies noted by HN almost exclusively have steep IR spectra,
while the non-knotty ones lie (mostly) in the lower luminosity
flat spectrum region, with a distribution among the steep spectrum sources
as well. The redshift distributions are comparable, so there should not be any
bias by resolution limits. In the spectrum class sequence (QSO, Sy1, Sy2,
LINER, H II, `galaxy'), which represents decreasing `activity', the knotty
galaxies are less active by more than one unit, and are
0.6 mag less luminous. The ages and strengths of the knotty galaxies are
younger and stronger by small amounts. The overall IR spectral energy
distribution from the four IRAS bands (12, 25, 60, 100$\mu$) is steeper
in the knotty galaxies. Galaxy colours are very comparable.

    Further inspection was carried out on the original images of 24 of the
galaxies
with redshift less than 0.030, to focus more closely on the bright knots.
{}From this, is became clear that there is no sharp distinction between
galaxies
with and without knots, but a range of knot `morphologies'. These were
quantified in a sequence of increasing knottiness as follows: 0 = smooth, with
no knots seen; 1 = a few large/bright knots seen outside the galaxy; 2 =
a few large/bright knots seen inside the galaxy; 2.5 = a few large/bright knots
both inside and outside the galaxy; 3 = many knots with a range of brightness
within the galaxy; 4 as 3, with knots outside the galaxy too. In this
classification, `few' typically means 5 - 30 knots, and `many' typically
is 100 to 200. This is a more careful classification of knots than given
in HN, but has a smaller sample due to lower redshift cutoff, and the
unavailability of some of the original images. Table 1 shows the statistics
of these objects, and Figure 1 shows them on the IR diagram.

   These new results support and extend the larger sample results from HN.
The most knotty objects have steep IR spectra or low IR luminosity. The
least knotty and the smooth galaxies have flat IR spectra. However, with
such small subset numbers, we do not have statistically significant numerical
differences, and there are many subjective quantities (knottiness,
age and strength of interaction, even spectroscopic type). In addition,
it appears that the most knotty galaxies have the lowest redshift, so
that we have not been completely successful in eliminating spatial resolution
bias. (On the other hand, degradation of the low redshift images to mimic
their appearance at 2-3 times the distance, does not change their knotty
classification, although it does lose the fainter knots.)

    It does seem to be a general conclusion that the knottiest galaxies
have less spectroscopically active nuclei. Also, the knottiest galaxies
are seen at an earlier stage of a tidal event: galaxies without knots
are seen at later stages.

   Table 1 includes galaxies of different types and luminosities.
However, there are several kinds of star-forming galaxies in the sample.
There are mergers, where a disturbed merger product galaxy is seen; there
are non-merging interactions where the companion is still seen; and there
are galaxies which do not appear to be tidally disturbed.
The latter are mostly lower luminosity galaxies, where
star-formation appears to be their undisturbed normal evolution.
In the case of non-merging encounters, we see
knots outside the main galaxy, presumably indicating star-formation in
gas in bridges between the galaxies, as discussed e.g. by Mirabel, Dottori, and
Lutz (1992).

    The radio observations were also inspected of the subsample: 15 of the 24
optically classified galaxies are in the data of Neff and Hutchings
(1992). Five of these are radio non-detections, and the others all slightly
(arcsec-level) extended nuclei with low flux. Of the ten detected sources,
seven have Seyfert or LINER spectra, consistent with an active radio nucleus.
The other 3 are classed as inactive galaxies but may have optically hidden
active
nuclei where the radio core sources are seen. None shows the ring-like radio
morphology seen in some IRAS galaxies, which is regarded as evidence
of circumnuclear star-formation. Thus, the star-formation knots we are studying
are not sources of significant radio emission from supernovae.

     This section has dealt with the circumstances in which bright knots
are seen. Assuming they indicate compact sites of star-formation, they
arise in irregular galaxies of no particular distinction, and also
are triggered by interactions. In the interacting cases, the knots arise
soon after the interaction, and evidently fade enough to become invisible
before the effects of the interaction have settled. They are also seen more
in galaxies with low nuclear activity: thus nuclear activity may indicate
cases where the gas is funneled to the galaxy centre rather than left
in the outer parts where it can condense into stars. Star-formation
knots outside the galaxy also occur, often in a non-merging encounter,
presumably where gas is drawn out of the galaxy by tidal forces, .

    The sample of galaxies is biased by being IR-bright, indicating
dust which may be heated by hot stars. The presence of dust itself is
associated with massive star-formation, and may obscure or alter the
optical flux from the knots. Thus, the IRAS sample is probably
different from galaxies where very blue or UV-bright knots are seen.
In the next section we look at photometry of the knots, to pursue
these points.
\vskip 10pt
\centerline{3. PHOTOMETRY}
\vskip 10pt
     Detailed photometry was performed on 6 galaxies, with a
range of knotty morphology. The galaxies chosen have a range of knotty
morphology and redshift but cover the range of IR luminosity and colour
of the lower panel of Figure 1 : in particular two of them are at redshift near
0.03, and so test the limits of resolution and flux. The photometry was done
on both B and R band images, observed as described in HN. The photometry was
done using DAOPHOT and ALLSTAR software (Stetson 1987; Stetson and Harris
1988),
and the PSF was defined using typical isolated knots in each image.
In what follows we discuss only knots that were measured in both colours.
In the highest redshift galaxy (0148+223) the knots are more irregular,
presumably because they are not separately resolved.
Their photometry was also done `manually'
on individual knots as a check on the automatic ALLSTAR results
(using the `rimexam' task of IRAF, which measures signal from apertures
centred on objects selected individually on the screen), and the agreement
was good (see Figures 2,3). Colour-magnitude diagrams were constructed for
the knots,
and these are shown in Figure 2. Conversions to absolute magnitude and B-V,
using H$_o$=100, and an empirical linear relation between B-V and B-R
are shown in Figure 3.
The formulae used are M$_{BOL}$ = (m$_B$+m$_R$)/2 + BC - DM,
where DM is the distance modulus derived from the redshift using H$_0$=100, and
BC is a color-dependent bolometric correction assuming
unreddened stars (e.g. Flower 1977). The color conversion used is
B-V = 0.54(B-R - 0.3),
also based on unreddened star values, and BC = -1.1 + 3.26(B-V) -
2.33(B-V)$^2$.
These conversions allow us to make comparisons with standard models and
have uncertainties comparable with that in DM by using the redshift
as distance indicator, and are of the order 0.3m. The basic photometric
calibrations
were taken from the observing run as reported by HN. Table 2 presents the
6 galaxies and the photometry results.

  As noted earlier, Figures 2 and 3 show that few of the knots
are very blue. Until we
obtain spectra we cannot be sure if they are reddened young stars or
evolved unreddened stars. In Table 2 we show the upper brightness
limits for the knots in each galaxy, ignoring single outliers which
may be blends or foreground objects - i.e. looking for an upper
limit of luminosity that may apply to the knot population. The reader
can judge these by inspection of Figure 3. Clearly, the lower brightness
limits of knots are strongly redshift-dependent, and in the lower redshift
galaxies we are presumably measuring the brightest individual stars.
The luminosity of the brightest knots is correlated with the distribution of
knots (the knot index in Table 2): they are brightest when there are few
of them within the galaxy
and faintest in the galaxies with many knots within them. However, there
is a range of $\sim$2 magnitudes for each type. There is also a systematic
change of mean or reddest colour of the brightest knots with the maximum
brightness
(Figure 4) suggesting that
reddening may be responsible for the spread. Indeed, the spread
is what would be expected for normal interstellar reddening.
If the knots have the same age and composition, they may have the same
intrinsic
colour.
Standard de-reddening of the reddest knots to a constant colour value does
reduce
the (1$\sigma$) scatter from 1.7 to 1.2 magnitudes. De-reddening also
strengthens the correlation between maximum luminosity and knot index,
as shown in Figure 4. If all knots have the intrinsice colour of hot stars then
the maximum bolometric luminosity is M$_{BOL}$-16 or less. However, evolution
of a population will also redden the knot with age.
\vskip 10pt
\centerline{4. NOTES ON INDIVIDUAL GALAXIES}
\vskip 10pt
    Figure 5 shows images of the six galaxies measured.
Since the redshifts, morphology, knot density and crowding are all different,
it is worth making a few individual comments on the galaxies.
Generally, the photometry had the purpose of measuring the main population
of unresolved knots in the image. Completeness depends strongly on the
local brightness and crowding. Artifacts such as cosmic ray events are largely
eliminated by the PSF-fitting of DAOPHOT, and  by discussing only those
measured
in both B and R. Most of the objects measured
appear to be real from visual inspection of the original and source-subtracted
images.

\bf 0110+006 (NGC 428) \rm This galaxy is large and irregular, with a very
knotty appearance throughout. Two bar-like bright regions suggest that
an interaction may be occuring, and one of the bars has more bright
knots than the other. The galaxy is not detected at radio wavelengths by NH.

   The reddest knots lie at the edges of the bars or in the outer parts of
the whole system. The bluest objects lie mostly in the bright bar but
also are concentrated in the second bar and in the outer arm. No knots have
the colour and magnitude of main sequence stars and dereddening places
tham all above the 100M$\odot$ line for single stars.

\bf 0140+134 (NGC 660) \rm This is a spectacular example of an interaction
with two dust lanes that dominate the optical appearance. The main galaxy looks
like a warped disk seen near to edge-on, and it is intersected by a strong
dust lane associated with a much less luminous disk. There are several
bright knots associated with the outer parts of the second dust lane,
as well as others lying around the main galaxy. The main galaxy has a
knotty appearance along its bright ridge.

    The reddest knots all lie outside the main galaxy, associated with
the strong dust lane, and also on the side opposite from the dust lane.
These are also the brightest of the knots. The population of outside knots
defines the upper layer of knot brightness at all colours. The faintest
objects measured trace the main galaxy. The bluest objects trace both
dust lanes.

    This is the lowest redshift galaxy in the group, and knots are measured
to faint magnitudes (but no fainter than in the highest redshift object).
The faintest objects have the luminosity of individual stars of mass as
low as 10M$\odot$, but are redder. Dereddening them corresponds to main
sequence 40M$\odot$ stars. The brightest knots are brighter than 100M$\odot$
stars, and dereddening would bring them up to M$_{BOL}$=-17 if they are
hot.

\bf 0148+223 (NGC 695) \rm This is the highest redshift object measured,
at some ten times the distance of the lowest redshift galaxies. Thus,
considerably less detail is resolved. The galaxy has the appearance of a
face-on spiral, with very knotty structure along the arms. NH do not
detect the galaxy as a radio source. Because the
knots are likely to be clusters of stars, and are just resolved in the images,
they were measured individually by Rimexam in IRAF, as well as by DAOPHOT,
using the few most isolated as the PSF. Figure 3 shows both sets of
measurement, which have similar distributions in the H-R diagram.

   These knots are generally much more luminous than in the lower redshift
galaxies, presumably because of lower spatial resolution. The distribution
is extended along the direction of the reddening line. The DAOPHOT-found
objects are heavily concentrated in the galaxy, with the few outliers
probably being background galaxies.

\bf0215+143 (NGC 877) \rm This galaxy is an inclined spiral, with dust
lanes and knots that are seen at all radii in the disk. It is probably
similar to 2159-321. Essentially no knots are seen outsdie the disk.
The bluest knots lie in the inner part of the galaxy, but none is very
blue, as in all objects in the sample.

\bf 2159-321 (NGC 7172) \rm This galaxy looks like a fat disk seen edge-on,
with a very wide and long dust lane. There are bright knots out of the plane
on one side. A large number of fainter objects was measured, presumably bacause
of the smooth and dark nature of the background. More faint objects are
found on the same side as the clearly visible bright ones,  and none of the
objects on the `poor' side is blue, which suggests
that the dust lane extends out of the galaxy, as in 0140+134.

   The distribution of knots that lie inside the galaxy,
define a redder and fainter population than the knots outside, as also seen
also in 0140+134. Dereddening to the main sequence yields luminosities
that all exceed 100M$\odot$ stars.

\bf 2325+085 (Mkn 533) \rm This is another higher redshift object, comparable
with 0148+223. The galaxy is a face-on spiral with asymmetrical arms, and
a tidal connection to a nearby compact elliptical galaxy. The elliptical
has a red `shell' on the side nearest to the main spiral. The galaxy has
a weak and slightly extended nuclear radio source. There are a few
bright knots inside the disk, and even fewer outside it. The background
is smooth and the knots do not have any blending.

   DAOPHOT found very few faint objects to measure in this galaxy, unlike
the other high redshift object (0148+223). The objects measured lie
within the distributions seen in the H-R diagrams of the other galaxies,
but only along the upper luminosity edge.
\vskip 10pt
\centerline{5. DISCUSSION}
\vskip 10pt
  The evidence from the IRAS-selected sample of galaxies is that star-formation
commonly occurs in bright knots soon after a triggering tidal event. The
phenomenon is found in objects with steep IR spectra, with a range of IR
luminosity, and a suggestion that steepness and spectral index are correlated
in the knotty galaxies (Figure 1). The galaxy
IR spectrum may well be determined by the rate and location of star-formation.
Star-formation knots have maximum observed luminosity of M$_{BOL}$ = -17.5
dereddened to a B-V of 1.0, while the observed maximum is -16 at
B-V of 1.4. From the H-R diagrams, it appears that dust alone cannot account
for the range in
maximum luminosity. Other possibilities are element abundances differences, or
age differences, or that upper mass limits range quite widely.

    It is of interest to see whether the objects observed here could evolve
into
present-day globular clusters. Such a cluster typically has M$_{BOL}$ = -8,
B-R = 1.2 and B-V = 0.7. Models of stellar populations made by the Gissel
program (Bruzual and Charlot 1993) with standard initial mass function
indicate that the B-V colour of a starburst protoglobular cluster should
be $\sim$0 after 1Gyr,
and 0.4 after 2 Gyr, evolving from bolometric magnitude -12 to -10.5 as it
does.
The initial luminosity may be a few magnitudes brighter.
We sketch this sequence in the plots of Figure 3. If star-formation
is triggered by the interactions we can still see, they must be less than
$\sim$1 Gyr old,
since the tidal distortions typically die away on this timescale or less.
The populations that we see are not this blue, although they are very luminous.
Thus, if they are young forms of present-day globular clusters, they must be
reddened by some tenths in B-V, even in the objects that lie far from the
obvious dust. As can be seen in
2159-321, even the knots outside the galaxy appear to be reddened.
This contrasts with the bluer knot distribution in NGC 1275 (Richer et al
1993),
and may indicate that these IR-selected objects have a widespread
distribution of dust.

Spectroscopy is clearly required to test these possibilities, and derive ages
and
abundances.
Spectra would help determine of there are age spreads among knots in each
galaxy, as well as age differences between galaxies that relate to time
since the triggering event for the star-formation. The timescale for tidal
interactions between galaxies indicate that the brightest knots are seen
well within 1 Gyr of the event: thus it appears that knot-formation occurs
early and does not continue for long.

   We are almost certainly observing a range of different phenomena in the
galaxies studied here. The fainter knots seen inside galaxies are individual
stars of mass $\sim20M_{\odot}$ and higher, and also clusters of stars, ranging
in luminosity and possibly not separately resolved in the higher redshift
galaxies. The knots outside the galaxies are generally brighter and are
less crowded. These may well be new clusters that evolve into an addition
to the population of globular clusters long after the triggering event.
It thus may be that the globular cluster population of a galaxy has a range
of ages and tells the merging and star-formation history of the galaxy.
Spectroscopy is planned to test a range of these objects for their
population and ages.
\vskip 30pt
\centerline{\bf References}

Bruzual G., and Charlot S., 1993, ApJ, 405, 538

Crabtree D. and Smecker-Hane T., 1994, BAAS, 26, 1499

Flower P.J. 1977, A+A, 54, 31

Holtzman J.A. et al, 1992, AJ, 103, 691

Hunter D., O'Connell R., Gallagher J.A., 1994, AJ, 108, 84

Hutchings J.B. and Neff S.G., 1991, AJ, 101, 434 (HN)

Hutchings J.B. et al 1991, ApJ, 377, L25

Mirabel F, Dottori H, Lutz D. 1992, A+A, 256, L19

Neff S.G. and Hutchings J.B. 1992, AJ, 103, 1746 (NH)

O'Connell R., Gallagher J.A., Hunter D., 1994, ApJ, 433, 65

Richer H., Crabtree D., Fabian A., Lin D., 1993 AJ, 105, 877

Stetson P.B. 1987, PASP, 99, 191

Stetson P.B. and Harris W.E. 1988, AJ, 96, 909
\vfill\eject
\centerline{\bf Captions to Figures}
\vskip10pt
1. The Infrared colour-magnitude diagram showing the properties of the samples
discussed. Top: whole HN sample, with objects of z$<$0.04  shown as filled
symbols. Centre: The z$<$0.04 subsample of HN where the + symbols show galaxies
classified as knotty. Lower: z$<$0.03 subsample with knottiness indicated as
follows - s = smooth; o = knots outside galaxy; i = knots inside galaxy; b =
knots
both inside and outside; M = galaxies with many ($>$100) knots. Dashed lines
divide the plane into the four regions of HN: knots are seen in galaxies
concentrated in region 3 (lower left).
\vskip10pt

2. Measured colour-magnitude diagrams for knots seen in both B and R in six
galaxies. Table 2 summarizes properties of the galaxies.
\vskip10pt

3. Photometry of knots converted to M$_{BOL}$/B-V plane. ZAMS is indicated
in corner of each panel, and average reddening vector.  Reddening is steeper
for blue stars and shallower for red stars. Approximate luminosities of
individual stars of 100M$_{\odot}$ and 40M$_{\odot}$ are shown for reference.
Also sketched is the unreddened integrated evolution path leading to a present
epoch globular cluster, with numbers in Gyr since initial starburst (see text).
Panel for 0148+223 shows `manually' measured points for comparison. Panel
for 2159-321 shows knots within galaxy as solid symbols.
\vskip10pt

4. Ensemble knot behaviour in galaxies measured. The brightest knot
luminosity may be related to the type of knot, and the red colour limit.
\vskip10pt

5. B-band images of the galaxies measured in detail, illustrating types
of knot and range of morphologies. The images have N down and E to the right,
and have long dimensions as follows: 0110+006, 0140+134, and 2159-321 - 215
arcsec;
0215+143 and 2325+085 - 107 arcsec; 0148+223 - 54 arcsec.

\vfill\eject

\centerline{Table 1.}
\centerline{Mean properties of galaxy groups}
\vskip 10pt
\baselineskip 15pt
\settabs 12\columns
\hrule
\vskip 5pt
\+Group &&&\# &z &type$^a$ &str$^b$ &age$^b$ &B-R &M$_R$ &L$_{60\mu}^c$
&log(25/60$\mu$)\cr
\hrule
\vskip 2pt
\hrule
\vskip 5pt
\+HN sample &&&65 &0.065 &2.4 &2.7 &2.6 &1.5 &-21.6 &10.9 &-0.52\cr
\+HN knotty z$<$0.04 &&&17 &0.014 &3.8 &2.8 &2.8 &1.7 &-20.4 &10.5 &-0.81\cr
\+HN other  z$<$0.04 &&&23 &0.020 &2.6 &2.4 &2.6 &1.8 &-21.0 &10.5 &-0.50\cr
\vskip 10pt
\centerline{-----------------}
\vskip 5pt
Subset of HN with z$<$0.029
\+All  &&&24 &0.015 &3.2 &1.8 &2.6 &1.8 &-20.6 &10.2 &-0.63\cr
\+No knots &&&6 &0.021 &2.5 &1.7 &1.7 &1.7 &-20.4 &10.0 &-0.30\cr
\+Few, outside galaxy &&&4 &0.015 &3.5 &1.5 &3.5 &2.1 &-21.2 &10.6 &-0.69\cr
\+Few, inside galaxy  &&&4 &0.019 &1.5 &2.8 &2.8 &2.0 &-21.9 &10.7 &-0.57\cr
\+Few, in- and outside &&&5 &0.010 &4.2 &2.4 &2.6 &1.5 &-19.7 &10.1 &-0.89\cr
\+Many knots &&&5 &0.007 &4.2 &1.0 &2.8 &1.8 &-20.0 &~9.8 &-0.75\cr
\vskip 3pt
\hrule
\vskip 8pt
$^a$ 0=QSO; 1=Sy1; 2=Sy2; 3=LINER; 4=H II; 5=`galaxy'

$^b$ strength of interaction increasing from 1 to 5; age of interaction
increasing in time from 1 to 5.

$^c$ Log in units of L$_{\odot}$

\vfill\eject

\baselineskip 16pt
\centerline{Table 2}
\centerline {Galaxies with detailed photometry}
\vskip 10pt
\settabs 11\columns
\hrule
\vskip 2pt
\hrule
\vskip 5pt
\+Name &&~~z &Index$^a$ &DM$^b$ &\#knots$^c$ &~~M$_{BOL}^d$ &&B-V
&&Max.lum$^e$\cr
\vskip 5pt
\hrule
\vskip 5pt
\+0110+006 &&0.004 &3 &30.5 &220 &-8 to -13 &&0.5 to 1.8 &&-12.5/1.8\cr
\+(NGC 428)\cr
\vskip 4pt
\+0140+134 &&0.003 &4 &29.8 &75 &-6 to -12.5 &&0.4 to 1.8 &&-11/1.8\cr
\+(NGC 660)\cr
\vskip 4pt
\+0148+223 &&0.033 &2 &35.1 &75 &-12 to -17.5 &&0.4 to 1.4 &&-16/1.4\cr
\+(NGC 695)\cr
\vskip 4pt
\+0215+143 &&0.013 &3 &33.0 &135 &-10 to -15 &&0.5 to 1.5 &&-14/1.0\cr
\+(NGC 877)\cr
\vskip 4pt
\+2159-321 &&0.009 &2.5 &32.2 &120 &-8 to -15.5 &&0.4 to 2.1 &&-13/1.9\cr
\+(NGC 7172)\cr
\vskip 4pt
\+2325+085 &&0.029 &2.5 &34.7 &15 &-13.5 to -16 &&0.6 to 1.5 &&-15/1.1\cr
\+(Mkn 533)\cr
\vskip 5pt
\hrule
\vskip 2pt
\hrule
\vskip 10pt
$^a$ Knot index: 1=few outside, 2=few inside, 2.5=few inside+outside,
3=many inside, 4=many inside+outside.

$^b$ Distance modulus in magnitudes for H$_0$=100, for galaxy redshift

$^c$ Number of knots measured in both B and R bands by DAOPHOT

$^d ~(m_B + m_R)/2 + BC - DM$  (see text)

$^e$ Absolute magnitude of bright limit of general population of knots,
and the red B-V limit of that population (see Figure 3)
\end